\documentclass[useAMS,usenatbib]{mn2e}

\usepackage{natbib}
\usepackage{graphicx}
\usepackage{amssymb}
\usepackage{color}

\title[AGN radiation pressure-driven outflows]{AGN feedback: galactic-scale outflows driven by radiation pressure on dust}
\author[ ]
{W. Ishibashi$^{1}$\thanks{E-mail:
wako.ishibashi@phys.ethz.ch} and A. C. Fabian$^{2}$
\footnotemark[0]\\
$^{1}$Institute for Astronomy, Department of Physics, ETH Zurich, Wolfgang-Pauli-Strasse 27, CH-8093 Zurich, Switzerland 
\footnotemark[0]\\
$^{2}$Institute of Astronomy, Madingley Road, Cambridge CB3 0HA, UK \\
}

\begin{document}

\pdfminorversion=4

\date{Accepted ? Received ?; in original form ? }

\pagerange{\pageref{firstpage}--\pageref{lastpage}} \pubyear{2012}

\maketitle

\label{firstpage}

\begin{abstract} 
Galaxy-scale outflows, which are thought to provide the link connecting the central black hole to its host galaxy, are now starting to be observed. However, the physical origin of the mechanism driving the observed outflows, whether due to energy-driving or radiation-driving, is still debated; and in some cases, it is not clear whether the central source is an active galactic nucleus (AGN) or a nuclear starburst. Here we study the role of radiation pressure on dust in driving galactic-scale AGN outflows, and analyse the dynamics of the outflowing shell as a function of the underlying physical parameters. 
We show that high-velocity outflows ($\gtrsim$1000 km/s) with large momentum flux ($\gtrsim 10 L/c$) can be obtained, by taking into account the effects of radiation trapping. In particular, the high observed values of the momentum boosts can be reproduced, provided that the shell is initially optically thick to the reprocessed infrared radiation. 
Alternatively, the inferred measurements of the momentum flux may be significantly biased by AGN variability. 
In this context, the observations of powerful outflows on kiloparsec scales, with no or weak signs of ongoing nuclear activity at the present time, could be re-interpreted as relics of past AGN episodes. 
\end{abstract}

\begin{keywords}
black hole physics - galaxies: active - galaxies: evolution  
\end{keywords}


\section{Introduction}

The energy and momentum released by accretion onto the central black hole are thought to couple to the surrounding medium through some form of active galactic nucleus (AGN) feedback. 
As the accretion energy clearly exceeds the binding energy of the galaxy bulge, AGN feedback must have a great impact on its host, and may explain the observed black hole-host galaxy correlations \citep[][and references therein]{Silk_Rees_1998, Fabian_1999, King_2003, Murray_et_2005, Fabian_2012}. 
Feedback from the central black hole is also invoked to account for a number of observed galaxy properties, in particular the quenching of star formation leading to the emergence of red and dead ellipticals \citep{Springel_et_2005, DiMatteo_et_2005, Croton_et_2006, Bower_et_2006}. 
In this context, AGN outflows are thought to provide the physical connection linking the small scales ($\lesssim$pc) of the central black hole to the large scales ($\gtrsim$kpc) of the host galaxy where star formation takes place. 

There is increasing observational evidence of galactic-scale outflows reported in the literature \citep{Fischer_et_2010, Feruglio_et_2010, Rupke_Veilleux_2011, Sturm_et_2011, Veilleux_et_2013, Cicone_et_2014}. 
These high-velocity, large-scale outflows are detected in different forms, including ionised, neutral, and molecular gas phases.
Molecular outflows may be of particular interest since the molecular gas represents the bulk of the interstellar medium from which new stars eventually form. 
The measured high mass outflow rates, which often exceed the star formation rates, also imply short gas depletion timescales \citep[e.g.][]{Sturm_et_2011, Cicone_et_2014}. This suggests that massive outflows can potentially remove gas from the entire galaxy, and thus critically affect the evolution of the host. Indeed, the detection of such powerful outflows has been posited as a proof of AGN feedback in action. 
The main question is what drives the observed outflows, and how the energy and momentum released by the central black hole actually couple to the interstellar medium of the host galaxy. 
In principle, AGN feedback can operate via a number of physical mechanisms, such as jets, winds, and radiation pressure  \citep[][and references therein]{Fabian_2012}. 
One way of driving large-scale outflows is via high-velocity winds that generate shockwaves propagating into the ambient medium \citep[e.g.][]{King_et_2011, Zubovas_King_2012, Faucher-Giguere_Quataert_2012}; another possibility is by direct radiation pressure on dust \citep[e.g.][see the latter for a discussion comparing the two driving mechanisms]{Fabian_1999, Murray_et_2005, Paper_1}. 
Focusing on radiative feedback, we note that the importance of radiation pressure on dust has already been highlighted in   a number of past works. 
The direct effects of radiation pressure on dusty gas were first considered and introduced in the context of AGN feedback by \citet{Fabian_1999}. \citet{Murray_et_2005} discussed the role of radiation pressure on dust in driving galactic-scale winds through momentum deposition; while \citet{Thompson_et_2005} examined the physics of starburst disks supported by radiation pressure on dust grains. 
Numerical simulations of radiative feedback, explicitly based on radiation pressure on dust, have also been carried out \citep{Debuhr_et_2011, Roth_et_2012, Krumholz_Thompson_2013, Davis_et_2014}. 
Most recently, \citet{Thompson_et_2015} studied the dynamics of radiation pressure-driven shells and subsequent applications to a wide variety of astrophysical sources. 

Here we wish to extend the work of \citet{Thompson_et_2015} by further investigating the role of radiation pressure on dust in driving galactic-scale outflows, and discuss its implications in the broader context of AGN feedback.
The paper is structured as follows. In Section \ref{Sect_Rad_Press}, we recall some generalities on radiation pressure on dust as a potential mechanism for driving large-scale outflows. In Section \ref{Sect_Fixed_shell}, we study the case of the fixed-mass shell, and analyse the dependence of the shell dynamics on the different physical parameters. 
We then consider the sweeping-up of ambient material (Section \ref{Sect_Expanding_shell}) and the possibility of AGN variability (Section \ref{Sect_AGN_variability}). 
We compare our model results with observations of AGN-driven outflows in Section \ref{Sect_Comparison}, and discuss the resulting implications in Section \ref{Sect_Discussion}.


\section{Radiation pressure on dust}
\label{Sect_Rad_Press}

The strength of the interaction between the AGN radiation and the surrounding medium depends on the underlying coupling process: electron scattering or dust absorption. 
It has been argued that transport by direct radiation pressure on electrons is problematic, as the galaxy becomes rapidly optically thin to the central radiation, and is indeed inadequate to drive large-scale outflows. 
In contrast, the coupling can be much stronger in the case of radiation pressure acting on dust grains, as the dust absorption cross section ($\sigma_d$) is much larger than the Thomson cross section ($\sigma_T$), with a ratio typically of the order of $\sigma_d/\sigma_T \sim 1000$ \citep[e.g.][]{Fabian_2012}. Correspondingly, the effective Eddington limit for dust differs from the standard Eddington limit by a factor $\sigma_d/\sigma_T$. 

A major fraction of the AGN bolometric luminosity is emitted in the blue bump component, peaking in the ultraviolet (UV) region. The UV photons are efficiently absorbed by dust grains embedded in the gas, the typical UV opacity being of the order of $\kappa_{UV}  \sim 10^3 \mathrm{cm^2 g^{-1}}$. 
As required by energy conservation, the absorbed UV emission is then reprocessed and re-emitted as infrared (IR) radiation. 
Note that the typical opacity of dust in the infrared ($\kappa_{IR}  \sim 5 \, \mathrm{cm^2 g^{-1}}$), although greater than the electron scattering opacity ($\kappa_{T} = \sigma_T/m_p \sim 0.4 \, \mathrm{cm^2 g^{-1}}$), is much smaller than at UV wavelengths \citep{Novak_et_2012}. 
The re-emitted IR photons can then either directly escape the medium without further interactions (if the medium is optically thin), or undergo multiple scatterings before diffusing out of the region (if the medium is optically thick). 

The momentum flux transferred from the radiation field to the gas is $L/c$ in the single scattering limit. 
Thus each absorption and scattering of UV radiation couples $L/c$ to the ambient medium. 
If the gas is optically thick to the re-radiated IR emission, additional momentum will be deposited due to the multiple scatterings: $\tau_{IR} L/c$ where $\tau_{IR}$ is the IR optical depth.


\subsection{Shell evolution}

We assume that radiation pressure drives a shell of dusty gas into the galaxy, and consider a thin shell geometry following \citet{Thompson_et_2015}. 
The general form of the equation of motion of the shell is given by:

\begin{equation}
\frac{d}{dt} (M_{sh}(r)v) = \frac{L}{c} (1 + \tau_{IR} - e^{-\tau_{UV}}) - \frac{G M(r) M_{sh}(r)}{r^2} 
\label{Eq_motion}
\end{equation} 
where $M(r)$ is the total mass distribution and $M_{sh}(r)$ is the shell mass. 
The total mass enclosed within a radius $r$ is modelled as an isothermal sphere: 

\begin{equation}
M(r) = \frac{2 \sigma^2 r}{G}  \;  (+ \frac{M_{sh}(r)}{2})
\end{equation}
where $\sigma$ is the velocity dispersion, and the additional term includes the self-gravity of the shell.
The driving term is provided by the central luminosity $L$; the IR and UV optical depths are respectively given by

\begin{equation}
\tau_{IR}(r) = \frac{\kappa_{IR} M_{sh}(r)}{4 \pi r^2} 
\end{equation}
\begin{equation}
\tau_{UV}(r) = \frac{\kappa_{UV} M_{sh}(r)}{4 \pi r^2} 
\end{equation} 
where $\kappa_{IR}$ = 5 $\mathrm{cm^2 g^{-1}}$ $f_{dg, MW}$ and $\kappa_{UV}$ = $10^3 \mathrm{cm^2 g^{-1}}$ $f_{dg, MW}$ are the IR and UV opacities, with the dust-to-gas ratio normalized to the Milky Way value. 

As discussed in e.g. \citet{Thompson_et_2015}, there are three main regimes of optical depth to be identified: optically thick to both UV and IR, optically thick to UV but optically thin to IR, and optically thin to UV. 
If the medium is optically thick to the reprocessed IR radiation, then most of the photons momentum can be transferred to the gas through radiative diffusion. Such conditions can be reached in the nuclear regions of dense starbursts and obscured AGN, typically on scales of $\sim$few hundred parsecs \citep{Thompson_et_2005}.
In less extreme environments, the re-emitted IR photons tend to leave the system without further interactions.  
In the single scattering regime (optically thick to UV but optically thin to IR), UV photons are absorbed, while the re-radiated IR photons freely escape. 
Such situations are met for a wide range of typical galaxies interstellar media.  
Further out, the ambient medium finally becomes optically thin to the central UV radiation. 
Due to their different dust opacities in the IR and UV regions, the IR transparency radius is always smaller than the UV transparency radius.


\section{Fixed-mass shell}
\label{Sect_Fixed_shell}

We first consider the case of a geometrically thin shell of constant mass ($M_{sh}(r) = M_{sh}$), starting from an initial radius $R_0$ with initial velocity $v_0$, and moving outwards in the galactic potential. 
In this simple case, analytical approximations can be obtained \citep[cf][]{Thompson_et_2015}, which allow us to gain some physical insight into the problem. In order to study how the evolution of the shell depends on the different physical parameters, we consider the variation of one single quantity at a time (while keeping the other values fixed), and analyse its effects on the shell dynamics. 

We assume the following values as fiducial parameters of the model: $\sigma$ = 200 km/s, $\kappa_{IR}$ = 5 $\mathrm{cm^2/g}$, $\kappa_{UV}$ = $10^3$ $\mathrm{cm^2/g}$, $R_0$ = 200 pc, $v_0$ = 0 km/s, $L = 5 \times 10^{46}$erg/s, $M_{sh} = 5 \times 10^{8} M_{\odot}$. 
For a shell of constant mass, the optical depth rapidly falls off with increasing radius, scaling as $\tau \propto 1/r^2$. 
The corresponding transparency radius for IR and UV radiation are respectively given by $R_{IR} = \sqrt{\frac{\kappa_{IR} M_{sh}}{4 \pi}} \sim 210 \kappa_{IR,0.7}^{1/2} M_{sh,8.7}^{1/2}$pc and $R_{UV} = \sqrt{\frac{\kappa_{UV} M_{sh}}{4 \pi}} \sim 3 \kappa_{UV,3}^{1/2} M_{sh,8.7}^{1/2}$kpc for fiducial parameters.
We note that the locations of the critical radii depend on the medium opacity and shell mass, scaling as $R_{IR,UV} \propto \kappa_{IR,UV}^{1/2} M_{sh}^{1/2}$.


\subsection{Radiative versus gravitational forces}

The evolution of the shell is governed by the balance between the inward force due to gravity and the outward force due to radiation pressure. 
The gravitational force is given by

\begin{equation}
F_{grav}(r) = \frac{G M(r) M_{sh}}{r^2} \propto \frac{1}{r}
\end{equation}
and roughly scales as $F_{grav} \propto 1/r$ for an isothermal mass distribution. 
The total radiative force is given by

\begin{equation}
F_{rad}(r) = \frac{L}{c} (1 + \tau_{IR} - e^{-\tau_{UV}})
\end{equation}

The shell is initially optically thick to the re-radiated IR emission up to $R_{IR}$, beyond which it becomes optically thin. At small radii, the radiative force is thus dominated by a combination of the infrared and single scattering terms
\begin{equation}
F_{rad}(r) \sim \frac{L}{c} (1 + \tau_{IR})
\end{equation}

At larger radii, where the shell becomes optically thin to the UV radiation, the radiative force is dominated by the ultraviolet term
\begin{equation}
F_{rad}(r) \sim \tau_{UV} \frac{L}{c} \propto \frac{1}{r^2} 
\end{equation} 
scaling as $1/r^2$, and we recover the classical optically thin limit.

\begin{figure}
\begin{center}
\includegraphics[angle=0,width=0.4\textwidth]{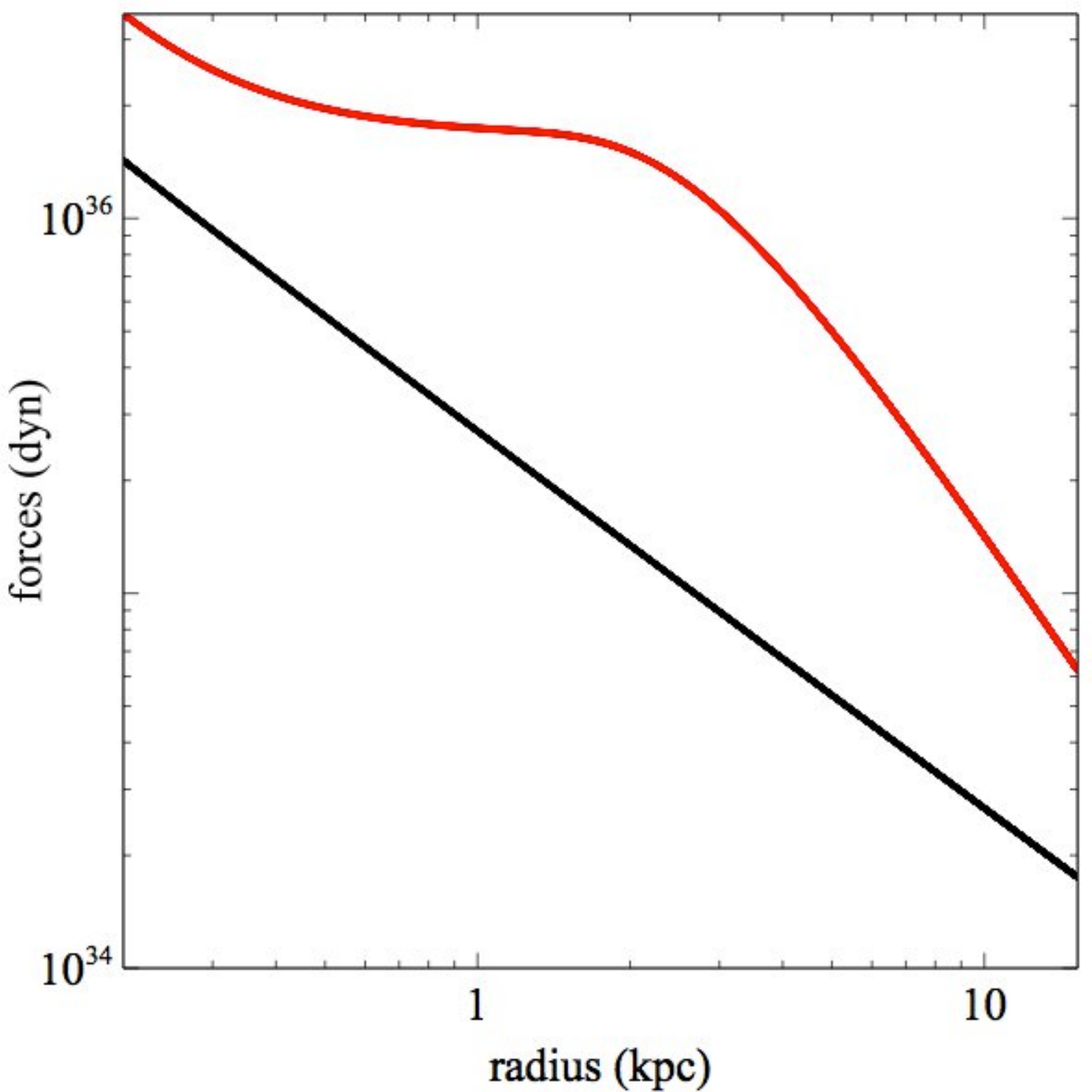} 
\caption{\small
Gravitational force (black) and radiative force (red) as a function of radius. 
}
\label{Fig_forces}
\end{center}
\end{figure} 

In Figure \ref{Fig_forces}, we plot the forces due to gravity (black curve) and radiation pressure (red curve) as a function of radius. 
We see that the radiative force exceeds the gravitational force for most radii, i.e. the ratio $F_{rad}/F_{grav}$ is greater than unity, implying that the shell is able to travel outwards. 
On large scales ($r \gtrsim R_{UV}$), the radiative flux decreases more rapidly with radius ($\propto 1/r^2$) than gravity ($\propto 1/r$), which eventually may dominate at very large radii. 


\begin{figure}
\begin{center} 
\includegraphics[angle=0,width=0.4\textwidth]{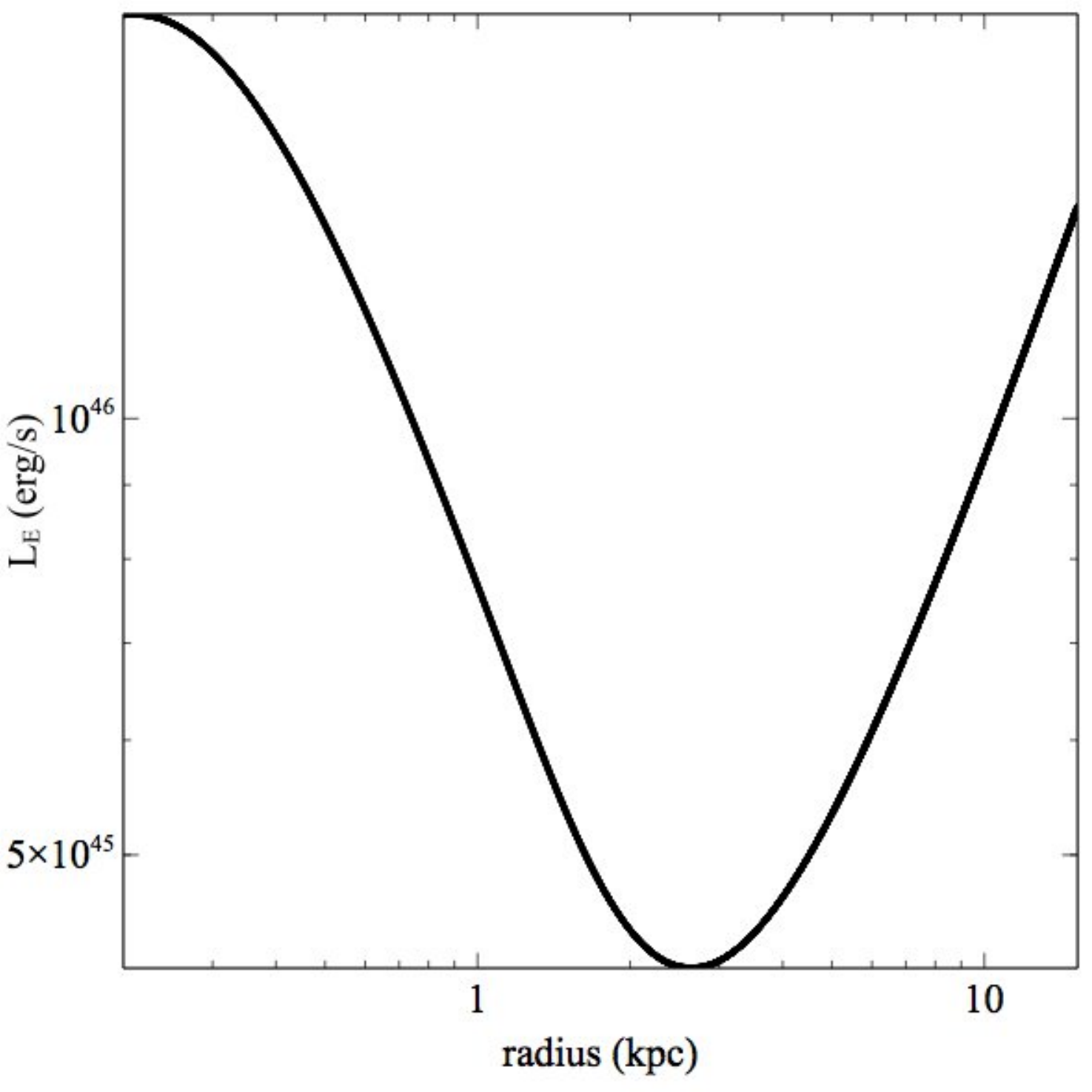} 
\caption{\small
Critical luminosity as a function of radius.  
}
\label{Fig_LEdd}
\end{center}
\end{figure} 

\subsubsection{Critical luminosity}

Equating the outward force due to radiation pressure to the inward force due to gravity, $F_{rad}(r) = F_{grav}(r)$, leads to a critical luminosity: 
\begin{equation}
L_E(r) = \frac{2 \sigma^2 M_{sh} c}{r} (1 + \tau_{IR} - e^{-\tau_{UV}})^{-1}
\end{equation}
which can be considered as a generalised form of the Eddington luminosity. 
The critical luminosity is plotted as a function of radius in Figure \ref{Fig_LEdd}. 
We observe that  $L_E(r)$ drops to a minimum around a few $\sim$kpc, corresponding to the location of the maximal gap between the radiative and gravitational forces (i.e. maximum of the $F_{rad}/F_{grav}$ ratio).


\subsection{Asymptotic velocity}

Neglecting the exponential term in the equation of motion (Eq. \ref{Eq_motion}), we obtain an analytical solution for the radial velocity profile \citep[cf Eq. 16 in][]{Thompson_et_2015}. 
The resulting asymptotic velocity can be approximated as:
\begin{equation}
v_{\infty} \sim 2 \sqrt{\frac{L}{c}} (\frac{\kappa_{UV}}{4 \pi M_{sh}})^{1/4} \propto \kappa_{UV}^{1/4} M_{sh}^{-1/4} L^{1/2} \\
\label{Eq_vinfty}
\end{equation} 

\begin{figure}
\begin{center}
\includegraphics[angle=0,width=0.4\textwidth]{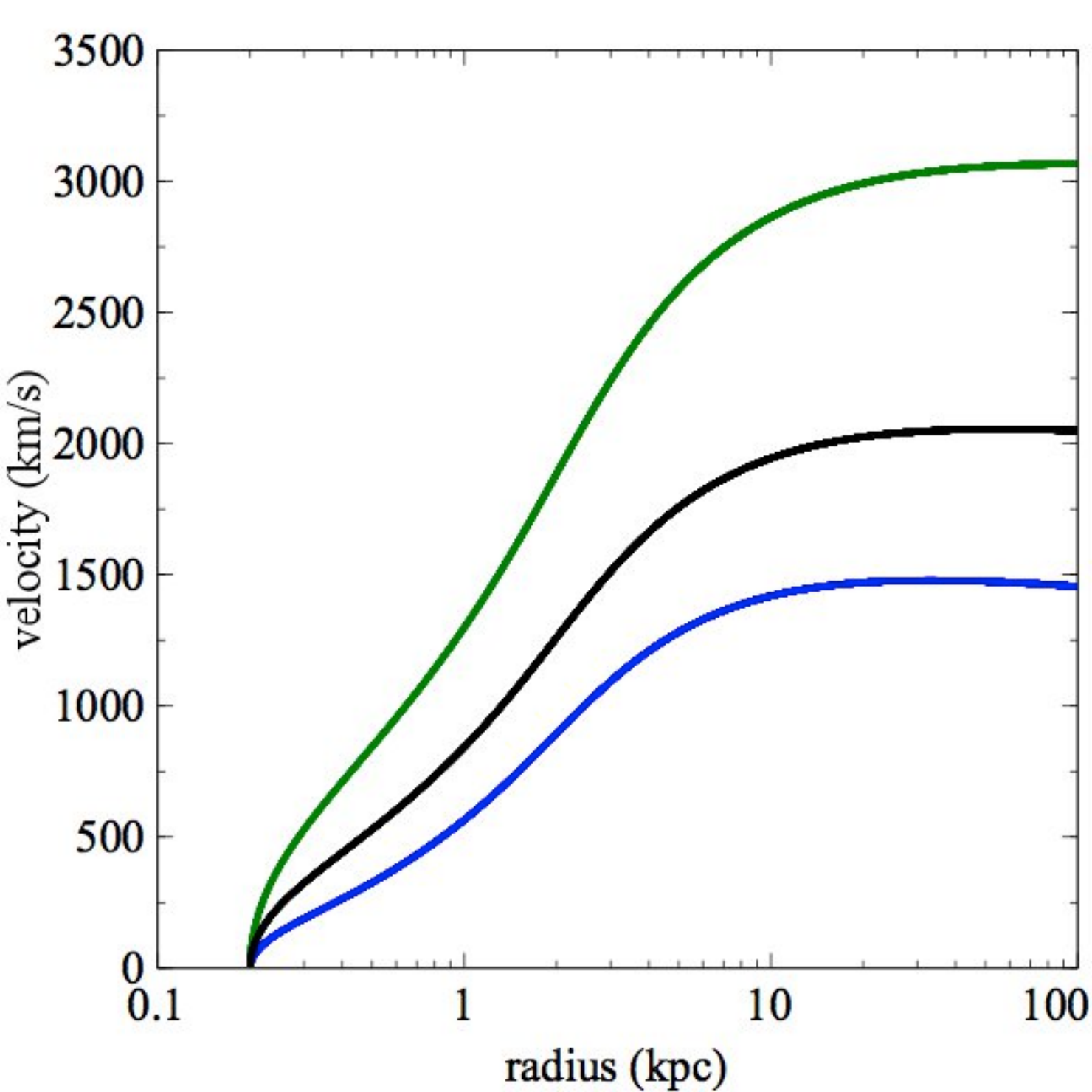} 
\caption{\small
Velocity as a function of radius for variations in luminosity: $L = 3 \times 10^{46}$erg/s (blue), $L = 5 \times 10^{46}$erg/s (black), $L = 10^{47}$erg/s (green).  
}
\label{Fig_varL}
\end{center}
\end{figure}

\begin{figure}
\begin{center}
\includegraphics[angle=0,width=0.4\textwidth]{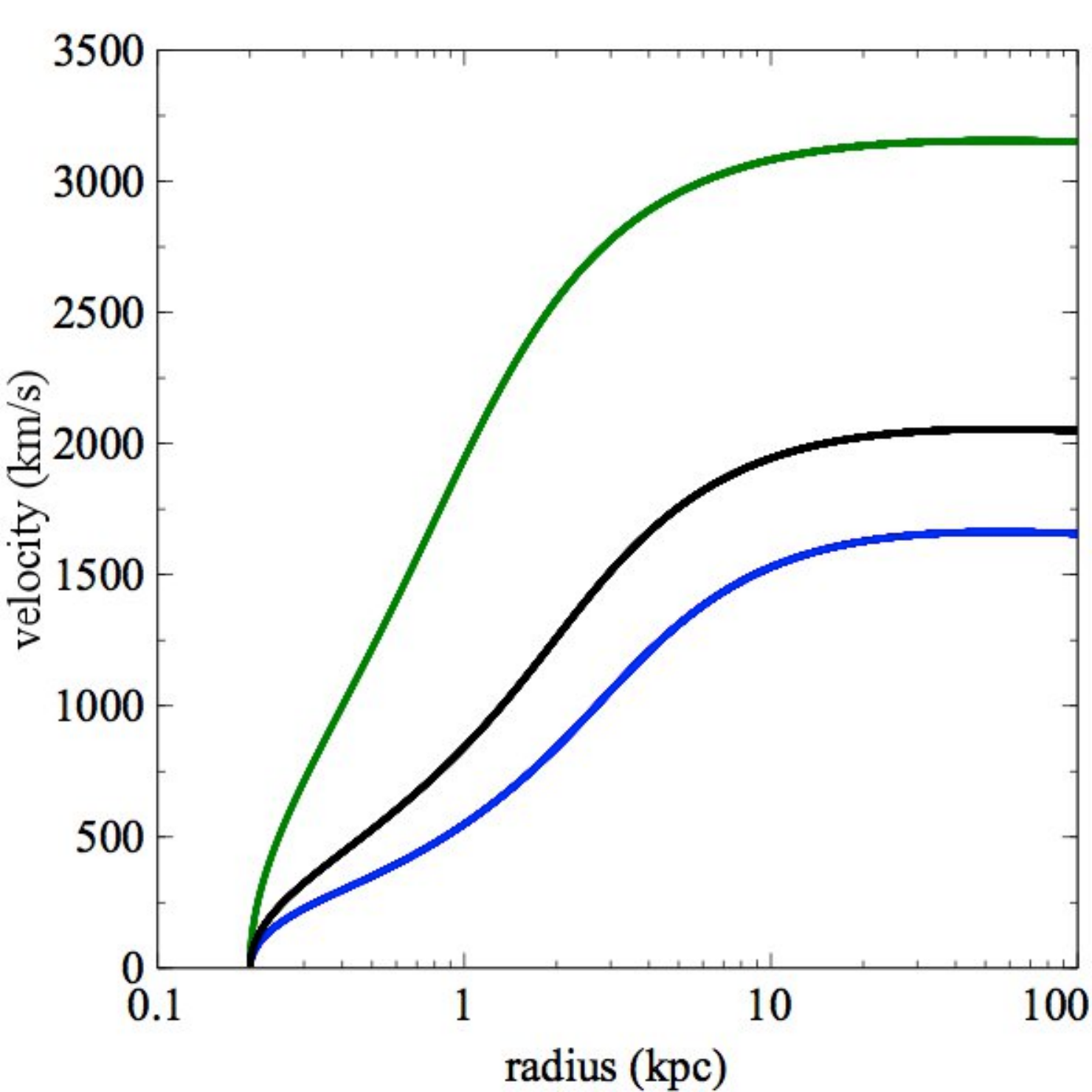}
\caption{\small
Velocity as a function of radius for variations in shell mass: $M_{sh} = 10^9 M_{\odot}$ (blue), $M_{sh} = 5 \times 10^8 M_{\odot}$ (black), $M_{sh} = 10^8 M_{\odot}$ (green). 
}
\label{Fig_varMsh}
\end{center}
\end{figure}

Numerically integrating the equation of motion, we obtain the exact radial velocity profile of the outflowing shell. 
We note that the isothermal potential approximation only holds in the inner regions (on $\sim$10 kpc-scales), but the results remain very similar assuming a NFW profile.  
Thus in the following, we show for simplicity the results obtained in the case of the isothermal potential.
Figures \ref{Fig_varL} and \ref{Fig_varMsh} show the velocity of the shell as a function of radius for different values of the central luminosity and shell mass, respectively. 
We see that the shell reaches higher velocities for higher luminosities, as expected. 
For a given luminosity, higher velocities are reached by lower mass shells. 
A variation in the central luminosity has a stronger effect on the velocity profile than a change in the shell mass. 
As can be seen from the analytical limit (Eq. \ref{Eq_vinfty}), we note that the dependence of the asymptotic velocity on central luminosity is stronger ($v \propto L^{1/2}$) than on shell mass ($v \propto M_{sh}^{-1/4}$).  
For a given central luminosity and shell mass, the asymptotic velocity also depends on the UV opacity of the medium. 
Larger values of $\kappa_{UV}$ lead to higher velocities, with considerable differences seen at large radii; an enhanced IR opacity also gives a small contribution to the increase in shell velocity. 

Considering variations in the shell initial conditions, we observe that the initial velocity has no significant effect on the shell dynamics, while a smaller initial radius leads to a slightly higher asymptotic velocity. 
Similarly, the shell attains higher velocities in systems with lower velocity dispersions, as might be expected in shallower gravitational potentials.


\subsection{Momentum ratio}

The momentum ratio of the shell is defined as

\begin{equation}
\zeta = \frac{M_{sh} v^2}{r \frac{L}{c}}
\label{Eq_momratio}
\end{equation} 

At large radii ($r \gtrsim R_{UV}$), the asymptotic momentum ratio is given by 

\begin{equation}
\zeta (r \gtrsim R_{UV}) \sim 2 \frac{R_{UV}}{r} \sim \sqrt{\frac{\kappa_{UV} M_{sh}}{\pi}} \frac{1}{r}
\label{Eq_zetaUV}
\end{equation} 
and scales as $1/r$. 

At small radii, close to the IR transparency radius ($r \sim R_{IR}$), the momentum ratio can be approximated as

\begin{equation}
\zeta (R_{IR}) \sim 2 \frac{R_{IR}}{R_0} \sim 2 \sqrt{\tau_{IR,0}} \sim \sqrt{\frac{\kappa_{IR} M_{sh}}{\pi R_0^2}} 
\label{Eq_zetaIR}
\end{equation} 

The maximal value of the momentum flux is attained at small radii ($r \lesssim R_{IR}$), where the shell is optically thick to the infrared radiation. 
If the initial radius lies close to the IR transparency radius ($R_0 \sim R_{IR}$), then the shell becomes rapidly optically thin to the re-radiated IR emission and $\zeta (R_{IR}) \sim 2$. 
We note that the momentum ratio can be significantly higher only if the optical depth to the re-radiated IR photons is much larger than unity at the launch radius. As we will see later, the IR optical depth due to radiation trapping is indeed the main parameter governing the momentum ratio of the outflowing shell. 

\begin{figure}
\begin{center}
\includegraphics[angle=0,width=0.4\textwidth]{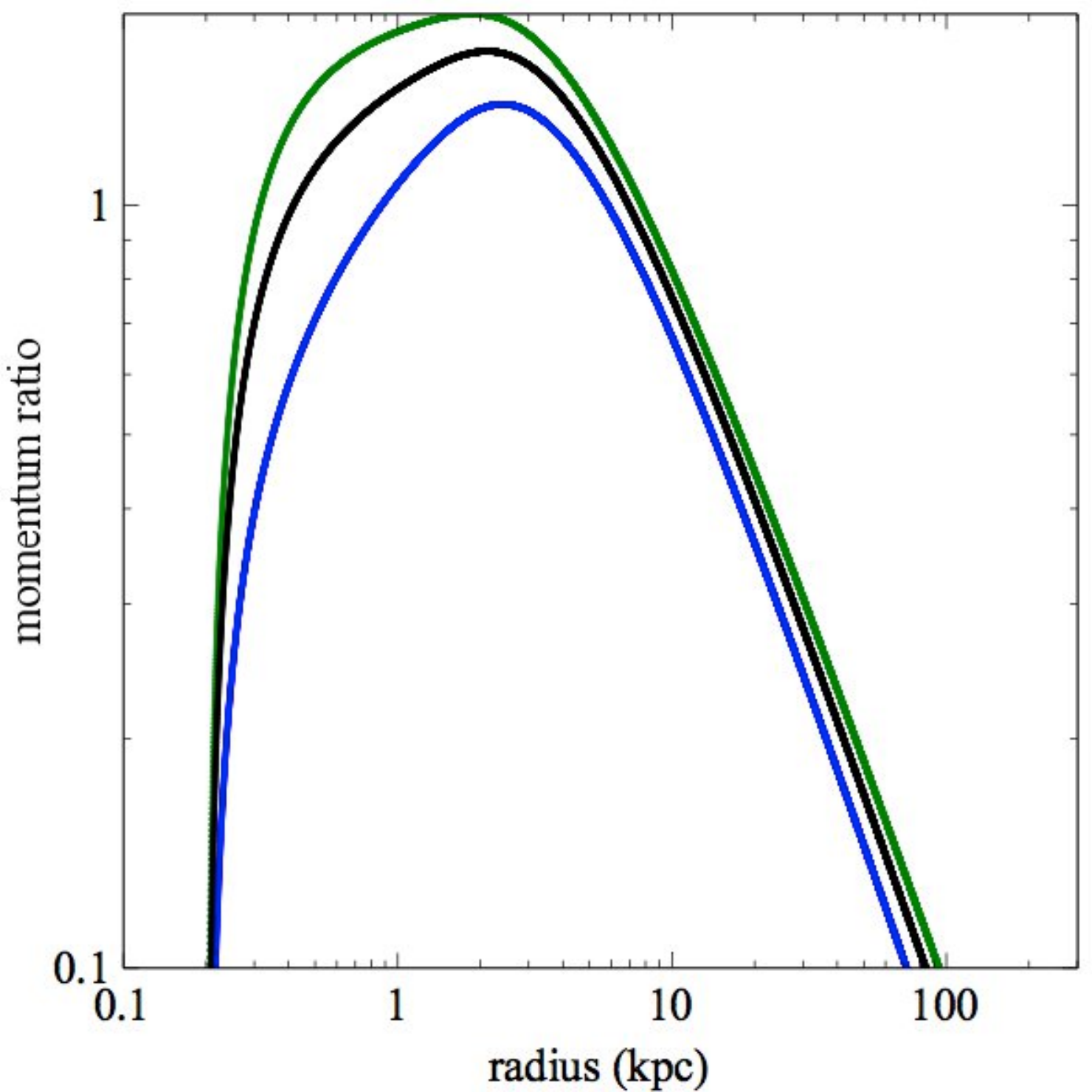}
\caption{\small
Momentum ratio as a function of radius for variations in luminosity: $L = 3 \times 10^{46}$erg/s (blue), $L = 5 \times 10^{46}$erg/s (black), $L = 10^{47}$erg/s (green).  
}
\label{Fig_momratio_varL}
\end{center}
\end{figure} 

Figure \ref{Fig_momratio_varL} shows the momentum ratio of the shell as a function of radius for different values of the central luminosity. 
We see that the momentum ratio always decreases as $1/r$ at large radii, while the peak values are roughly of the same order ($\zeta_{max} \sim 1$). 
A qualitatively similar plot is obtained in the case of varying shell mass. 
Thus, in contrast to the previously discussed velocity profiles, the momentum ratio is not very sensitive to variations in central luminosity and/or shell mass. 

\begin{figure}
\begin{center}
\includegraphics[angle=0,width=0.4\textwidth]{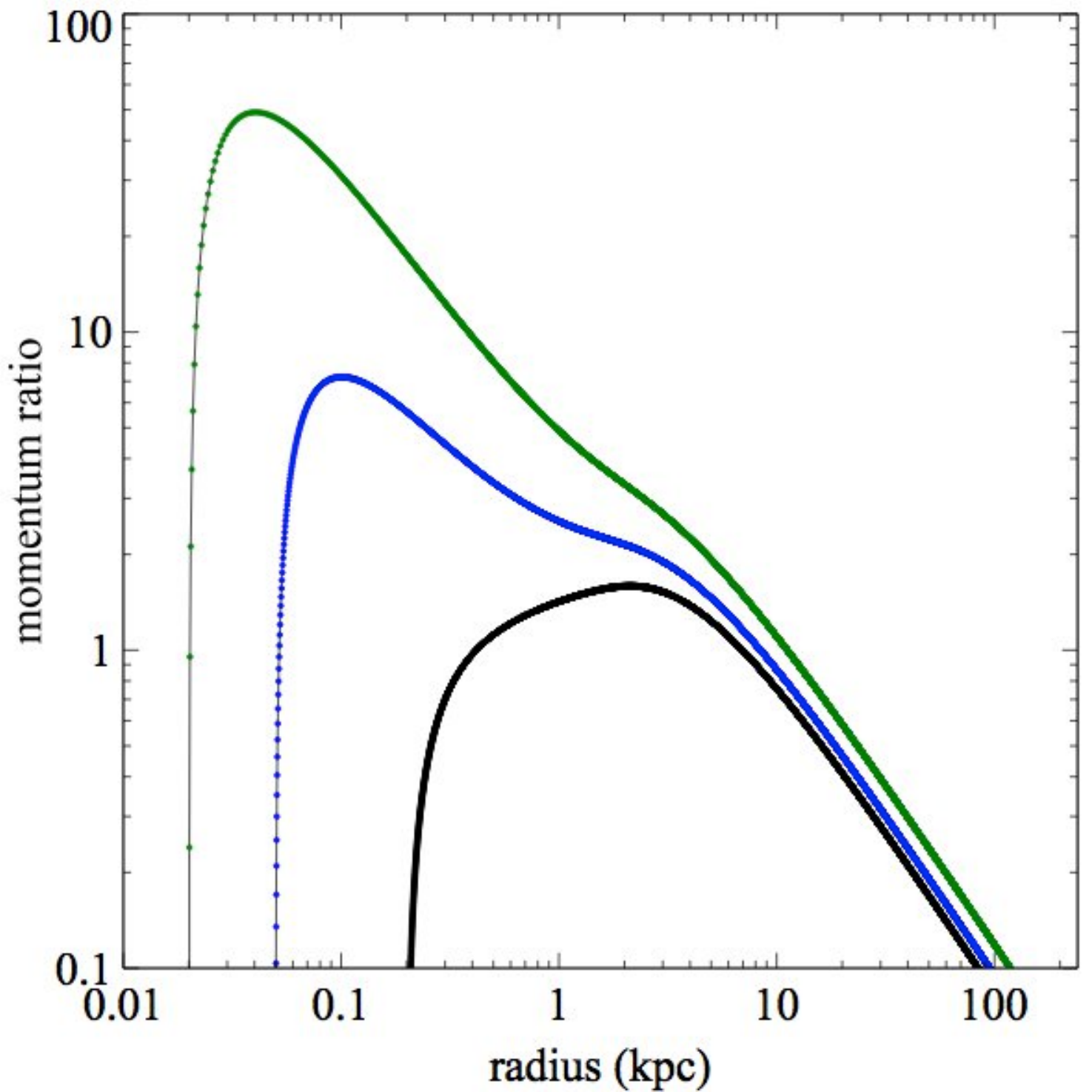} 
\caption{\small
Momentum ratio as a function of radius for variations in initial radius: 
$R_0$ = 200 pc (black), $R_0$ = 50 pc (blue), $R_0$ = 20 pc (green). 
}
\label{Fig_momratio_varR0}
\end{center}
\end{figure}

On the other hand, variations in the initial IR optical depth can have a significant impact on the momentum flux of the outflowing shell, with a larger initial IR optical depth leading to significantly higher $\zeta$ values. 
In physical terms, the increase in the initial IR optical depth can be due either to an enhanced IR opacity or a smaller initial radius. 
An enhanced opacity can be obtained e.g. by assuming a higher dust-to-gas ratio, as might be the case in the central regions of dense ULIRGs. 
In Figure \ref{Fig_momratio_varR0} we plot the radial profile of the momentum ratio for 3 different values of $R_0$. 
We see that larger momentum ratios are obtained for smaller initial radii. We also note that high $\zeta$ values are only observed on small scales ($\lesssim$kpc); while at larger radii ($\gtrsim$kpc), the momentum flux profiles rapidly fall off as $\propto 1/r$. 
From Eq. \ref{Eq_zetaIR}, we note that $\zeta(R_{IR})$ scales inversely with the initial radius $R_0$. 
A lower limit on the initial radius is set by the dust sublimation radius given by
\begin{equation}
R_{sub} = \sqrt{\frac{L}{16 \pi \sigma_{SB} T_{sub}^4}}
\end{equation}
where $T_{sub}$ is the sublimation temperature of the dust. 
For typical values ($T_{sub}$=1500K), the sublimation radius is of the order of $R_{sub} \sim 1$pc.

We also verify that variations in the other parameters do not affect the momentum ratio of the shell in a significant way. 
In particular, changes in the UV opacity do not have an appreciable effect on the value of the momentum flux. 

Summarising, a favourable combination of the different physical parameters can lead to high velocity outflows with large momentum fluxes. 
For a given luminosity and shell mass, the asymptotic velocity of the shell is mainly determined by the UV opacity; whereas the momentum flux (and in particular its peak value) is most sensitive to variations in the IR optical depth. 
On the other hand, we note that changes in the UV opacity have no significant impact on the momentum ratio; conversely, variations in the IR optical depth do not much affect the asymptotic velocity of the shell. 
These results are in qualitative agreement with the analytical limits given in Eqs. \ref{Eq_zetaUV} and \ref{Eq_zetaIR}. 
We can thus disentangle the respective contributions of the UV and IR terms to the overall dynamics of the outflowing shell.  


\subsection{Outflow rate}

The mass outflow rate in the shell limit can be defined as: 
\begin{equation}
\dot{M} \sim \frac{M_{sh}}{\tau} \sim \frac{M_{sh} v}{r}
\end{equation}

\begin{figure}
\begin{center} 
\includegraphics[angle=0,width=0.4\textwidth]{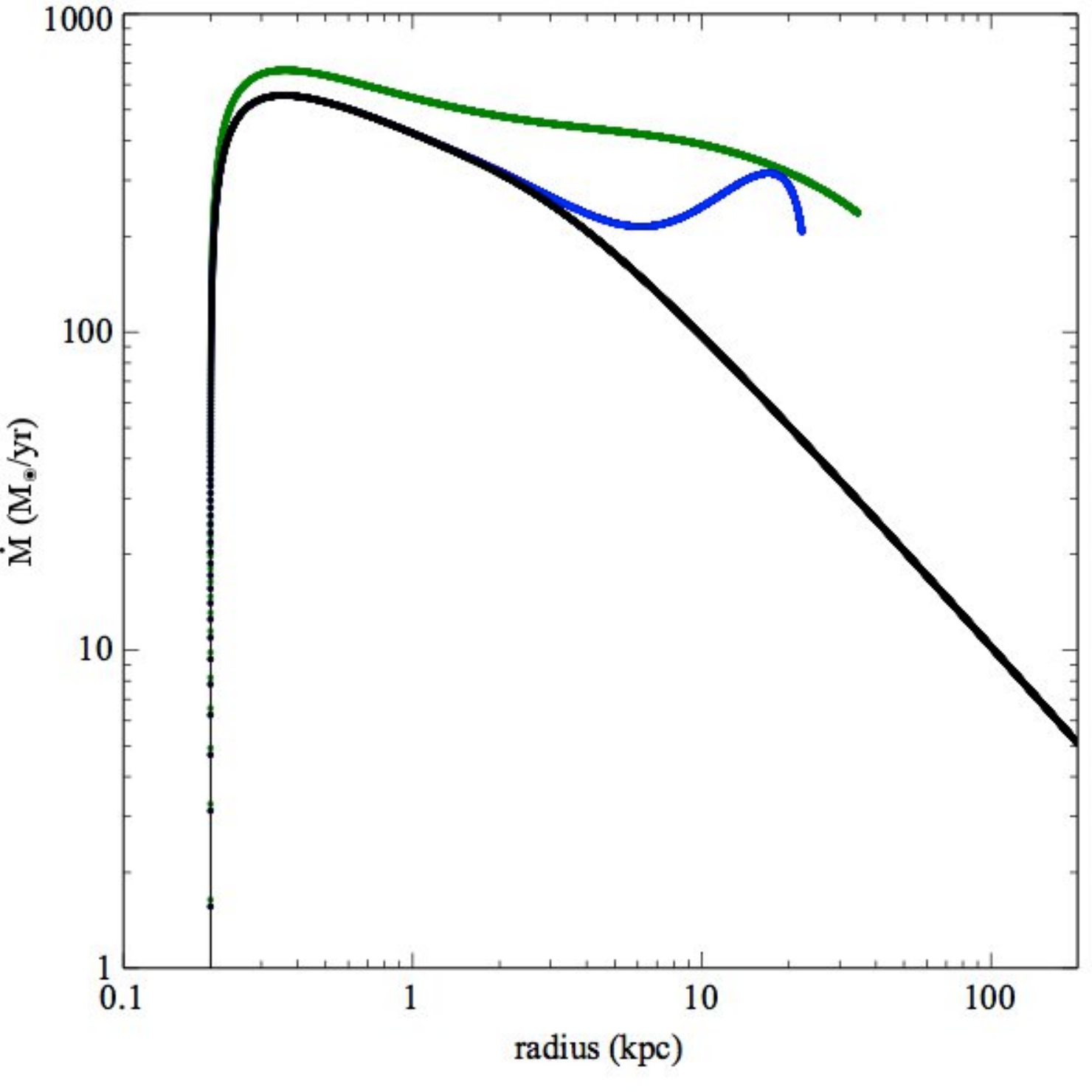} 
\caption{\small
Mass outflow rate as a function of radius for different shell models: fixed mass shell with $M_{sh} = 5 \times 10^8 M_{\odot}$ (black), shell sweeping up mass from a constant density medium with $n_0 = 0.1 cm^{-3}$ (blue), shell sweeping up mass from an isothermal density distribution with $f_g = 0.05$ (green). }
\label{Fig_outflowrates}
\end{center}
\end{figure}

In Fig. \ref{Fig_outflowrates}, we plot the mass outflow rates as a function of radius for different shell models (cf Sect. \ref{Sect_Expanding_shell}). In the case of a fixed mass shell, the outflow rate can reach values of several hundred solar masses per year, followed by a decline scaling as $\dot{M} \propto 1/r$ (black curve).

\section{Expanding shells}
\label{Sect_Expanding_shell}

In more realistic situations, the shell is likely to sweep up mass as it expands into the interstellar medium of the host galaxy. The previous analysis of the dynamics of fixed-mass shells may be a good approximation in cases where 
the subsequently added swept-up mass ($M_{sw}(r)$) is negligible compared to the initial mass of the shell ($M_{sh,0}$). 
But when the former becomes comparable to or exceed the initial mass, one has to explicitly take into account the contribution of the swept-up material. 
More specifically, the mass of the swept-up material on scales of the UV transparency radius ($M_{sh}(R_{UV}$)) should be compared with the initial mass of the shell (as discussed in \citet{Thompson_et_2015}).

The density distribution of the ambient medium can be parametrized as a power law of radius with slope $\alpha$:
\begin{equation}
n(r) = n_0 \left( \frac{r}{R_0} \right)^{-\alpha}
\end{equation}
where $n_0$ is the density of the external medium.
The corresponding swept-up mass is given by:
\begin{equation}
M_{sw}(r) = 4 \pi m_p \int n(r) r^2 dr = 4 \pi m_p n_0 R_0^{\alpha} \frac{r^{3-\alpha}}{3-\alpha}
\end{equation}

In the following, we focus on two particular cases of the ambient density profiles: 
$\alpha = 0$ and $\alpha = 2$, corresponding to constant density and isothermal distributions, respectively. 
The dust-to-gas ratio of the swept-up matter is parametrized as $\xi = f_{dg}/f_{dg, MW}$, normalized to the Milky Way value. 


\subsection{Constant density distribution}

If the ambient medium is characterized by a constant density distribution ($\alpha = 0$), the corresponding swept-up mass is given by $M_{sw,cst}(r) = \frac{4}{3} \pi m_p n_0 r^3 \cong 2 \times 10^7 n_{0.2} r_{kpc}^3 M_{\odot}$. 
The critical density for which the swept-up mass on scales of $R_{UV}$ is equal to the initial shell mass is: $n_{c} \cong 0.2 \kappa_{UV,3}^{-3/2} M_{sh,8.7}^{-1/2} \mathrm{cm^{-3}}$. 
As the swept-up mass increases very rapidly with radius (due to the steep radial dependence of the form $M_{sw,cst}(r) \propto r^3$), it quickly becomes dominant at large radii, even for external densities lower than the critical value (cf \citet{Thompson_et_2015}).
As in the case of the fixed-mass shell, variations in the physical parameters lead to qualitatively similar effects on the shell dynamics. 
For instance, outflowing shells reach higher velocities with increasing luminosity and decreasing mass. 

\begin{figure}
\begin{center}
\includegraphics[angle=0,width=0.4\textwidth]{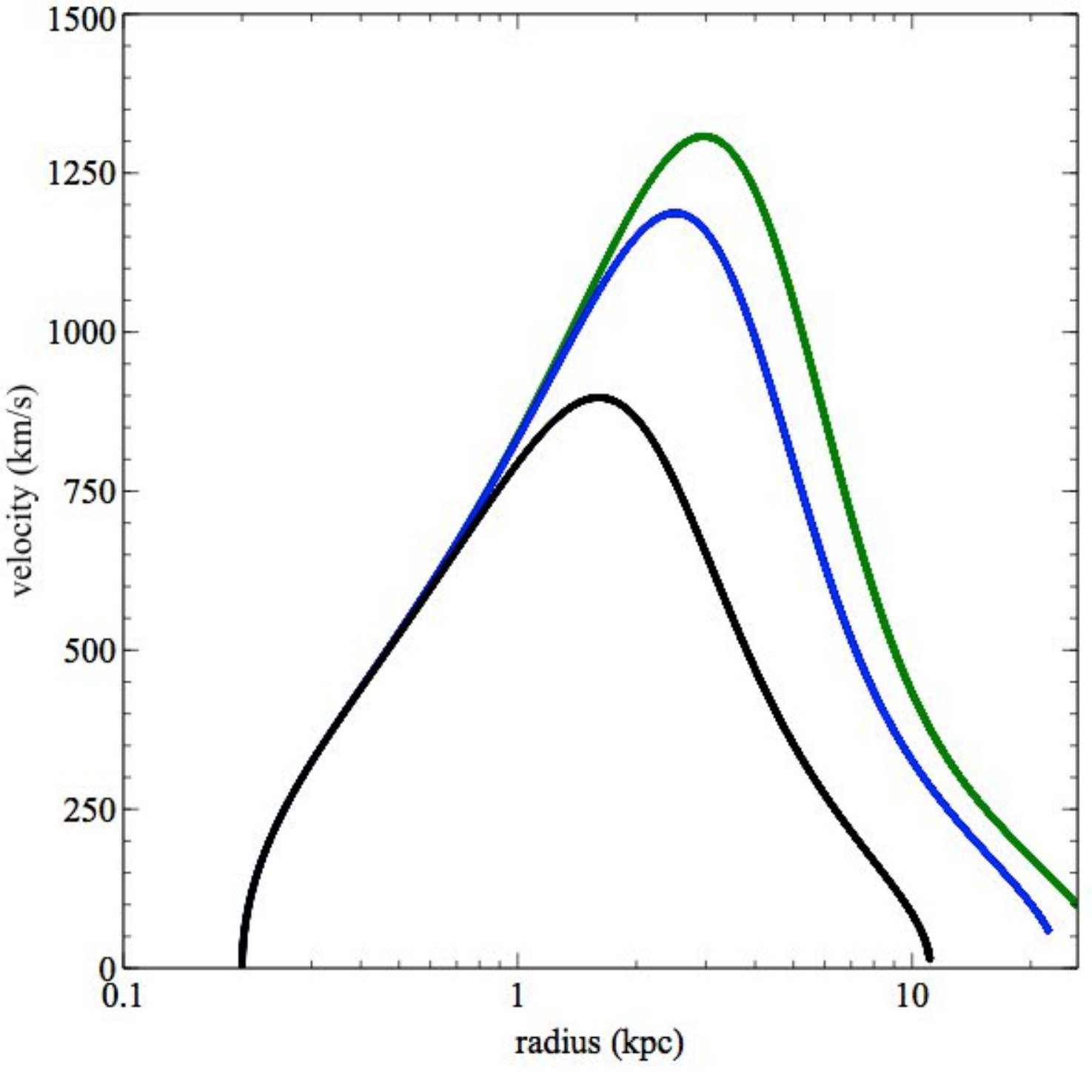}
\caption{\small
Velocity as a function of radius for various values of spatially constant density: $n_0 = 0.5 cm^{-3}$ (black), $n_0 = 0.1 cm^{-3}$ (blue), $n_0 = 0.05 cm^{-3}$ (green). 
}
\label{Fig_v_r_varNext}
\end{center}
\end{figure}

Here we focus on variations in the value of the external density ($n_0$). 
From the plot of the radial velocity profile (Fig. \ref{Fig_v_r_varNext}), we see that the shell reaches a certain maximal speed and then slows down, due to the sweeping up of ambient material. 
At small radii, the swept-up mass is still negligible compared to the initial mass of the shell, and the v(r) curves are roughly identical. 
In fact, the $M_{sh}(r)/M_{sh,0}$ mass ratio starts to exceed unity at relatively large radii, around $r_c \cong 3 n_{0.2}^{-1/3} M_{sh,8.7}^{1/3}$kpc, on scales comparable to $R_{UV}$.

\begin{figure}
\begin{center}
\includegraphics[angle=0,width=0.4\textwidth]{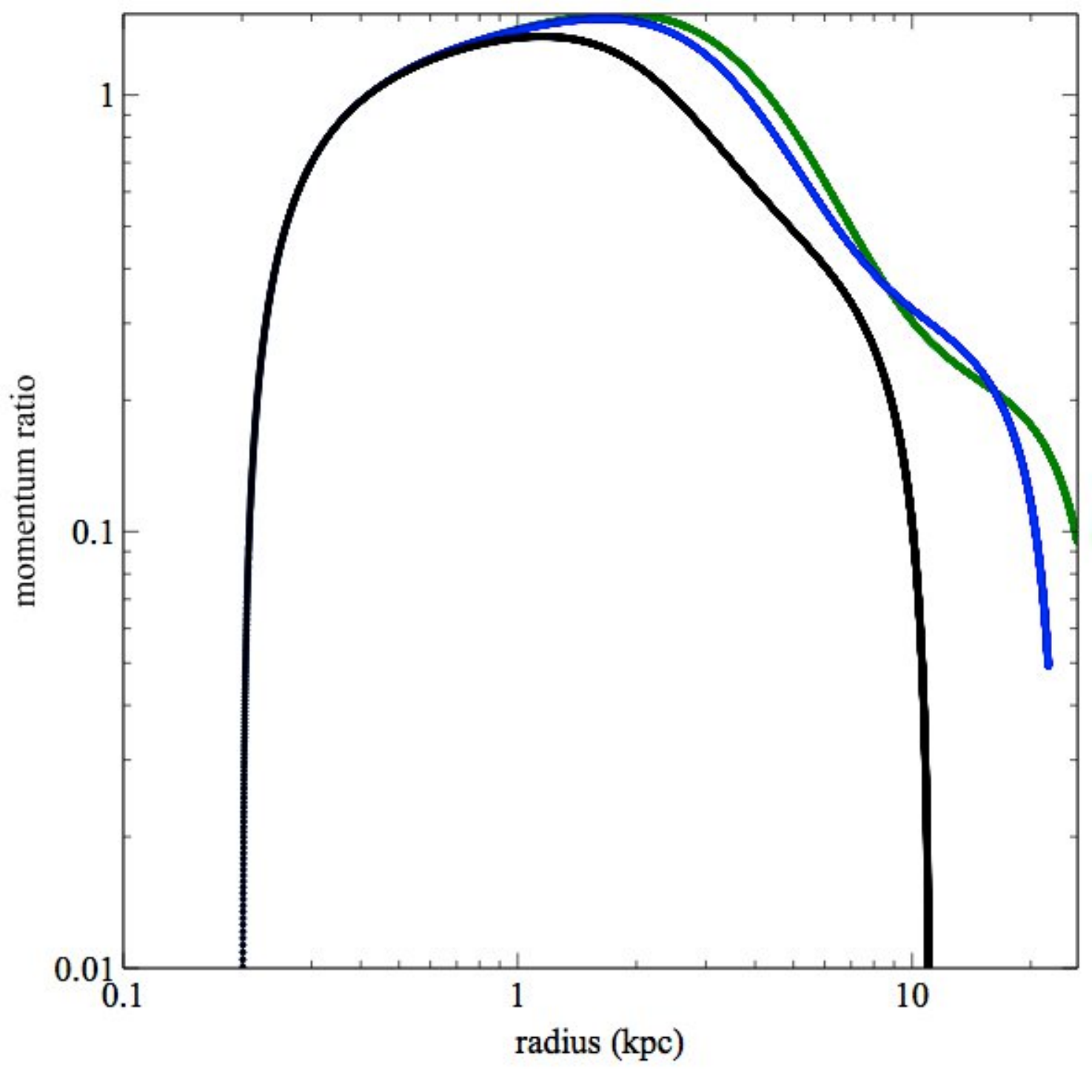}
\caption{\small
Momentum ratio as a function of radius for various values of spatially constant density: $n_0 = 0.5 cm^{-3}$ (black), $n_0 = 0.1 cm^{-3}$ (blue), $n_0 = 0.05 cm^{-3}$ (green). 
}
\label{Fig_momratio_r_varNext}
\end{center}
\end{figure}

The peak velocity attained and the subsequent deceleration pattern depend on the density of the surrounding medium.  
As the external density increases, the shell sweeps up more mass and attains a lower peak velocity followed by a rapid deceleration. 
Since the velocity rapidly falls off at large radii, it is unlikely that the shell is able to escape the galactic halo, and plausibly falls back. 
On the other hand, we observe that variations in the values of the external density do not have a dramatic effect on the momentum ratio of the shell (Fig. \ref{Fig_momratio_r_varNext}). 

In Fig. \ref{Fig_outflowrates}, we also plot the outflow rate for a shell sweeping up material from the surrounding medium of constant density (blue curve). 
As expected, the outflow rate scales as $\dot{M} \propto r^2 v$ at large radii, where the swept-up mass dominates the shell dynamics.

\subsection{Isothermal density distribution}

If the ambient density profile follows an isothermal distribution ($\alpha = 2$), the swept-mass is given by $M_{sw,iso}(r) \sim \frac{2 f_g \sigma^2}{G} r \cong 2 \times 10^8 f_{g,0.01} \sigma_{200}^2 r_{kpc} M_{\odot}$, where $f_g$ is the gas fraction. The critical value of the gas fraction for which the swept-up mass at $R_{UV}$ equals the initial mass is: $f_c \cong 0.01 \kappa_{UV,3}^{-1/2} \sigma_{200}^{-2} M_{sh,8.7}^{1/2}$.
Again, varying the different physical parameters lead to qualitatively similar outcomes to the case of fixed-mass shells.    

\begin{figure}
\begin{center}
\includegraphics[angle=0,width=0.4\textwidth]{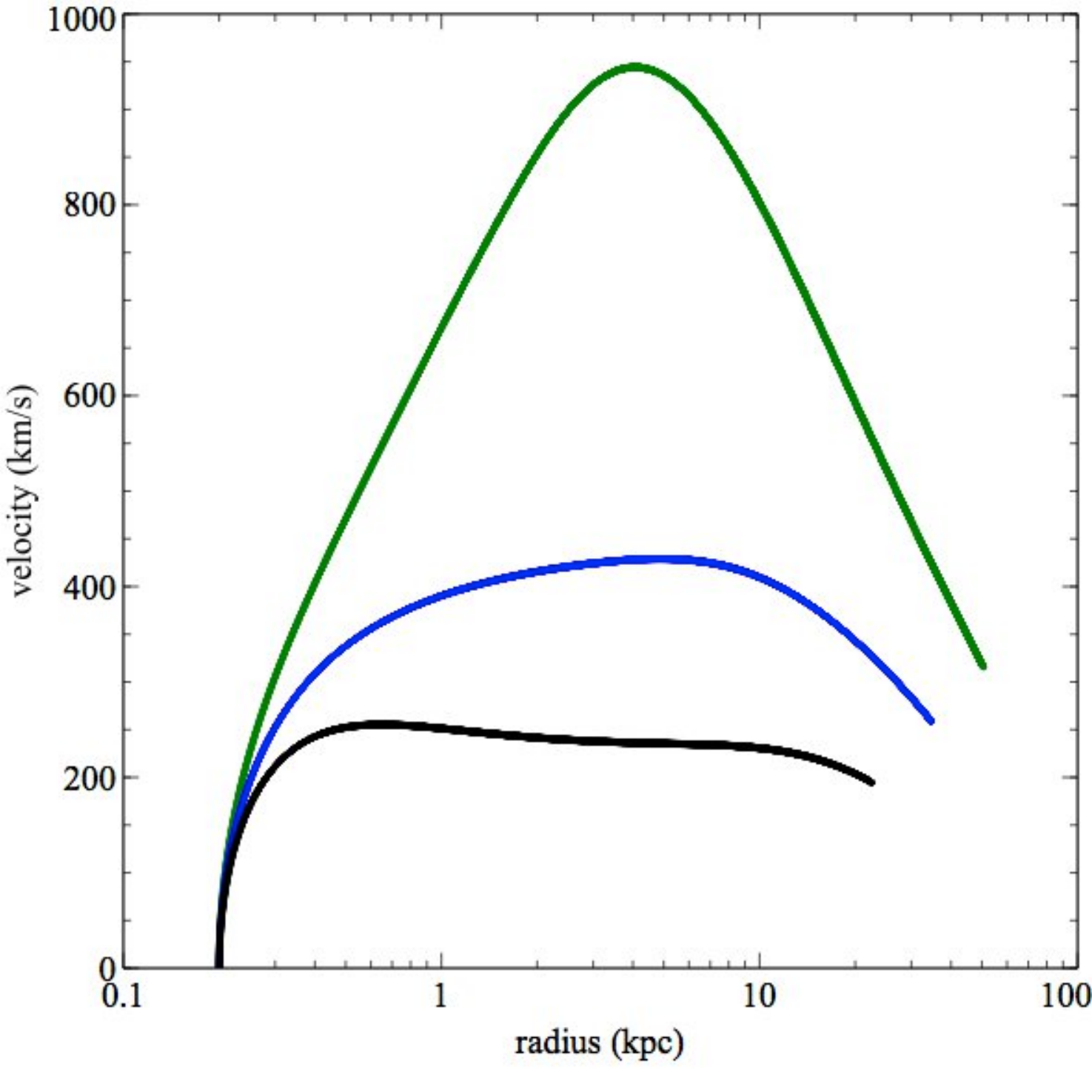} 
\caption{\small
Velocity as a function of radius for various values of the gas fraction: $f_g = 0.1$ (black), $f_g = 0.05$ (blue), $f_g = 0.01$ (green).
}
\label{Fig_v_r_varFg}
\end{center}
\end{figure}

\begin{figure}
\begin{center}
\includegraphics[angle=0,width=0.4\textwidth]{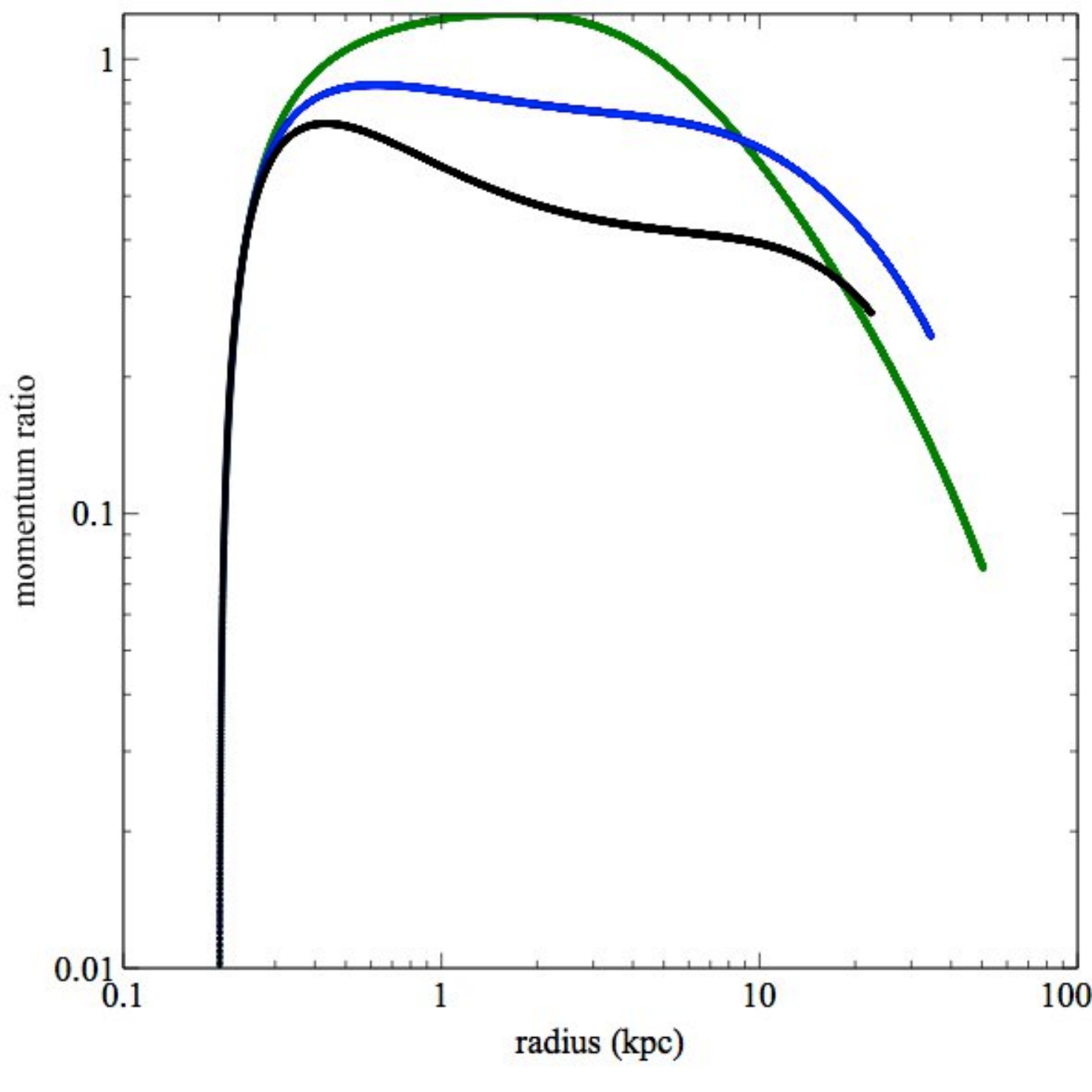} 
\caption{\small
Momentum ratio as a function of radius for various values of the gas fraction: $f_g = 0.1$ (black), $f_g = 0.05$ (blue), $f_g = 0.01$ (green).
}
\label{Fig_momratio_r_iso}
\end{center}
\end{figure}

We concentrate here on variations in the gas fraction, which is the most important parameter in determining the shell dynamics. 
As expected, the shell velocity is lower for ambient medium having higher gas fractions (Fig. \ref{Fig_v_r_varFg}). 
For $f_g > 0.05$, the shell can never escape the galaxy as its velocity is always lower than the escape velocity ($v < 2 \sigma$). For a smaller gas fraction ($f_g = 0.01$), the shell reaches a higher peak velocity temporarily exceeding the escape speed, but later falls back. This suggests that the outflow is bound on large scales, although the actual escape condition may depend on the large-scale gravitational potential. 
In contrast to the velocity profiles, we note that the momentum ratio is not significantly affected by variations in the gas fraction (Fig. \ref{Fig_momratio_r_iso}).
Compared to the constant density case, the swept-up mass in the isothermal density case rises more gently with radius, following a linear relation of the form $M_{sw,iso}(r) \propto r$. 
The amount of swept-up material in the two cases are comparable over a limited range of radii around $R_{UV}$, for the spanned external densities and gas fractions. The swept-up masses for the constant and isothermal density cases quickly diverge at larger radii, due to their distinct radial dependence.
In Fig. \ref{Fig_outflowrates}, we also plot the outflow rate for a shell sweeping up matter from the surrounding environment with an isothermal density distribution (green curve). We observe that the mass outflow rate is proportional to the velocity ($\dot{M} \propto v$), and is of the order of a few hundred solar masses per year.

The decrease of the gas mass fraction over cosmic time has been reported in a number of works \citep{Daddi_et_2010, Tacconi_et_2010}. High-redshift galaxies tend to have gas fractions much higher than the values observed in present-day galaxies. The depletion of gas could be partly due to past outflow events, which may lead to lower gas fractions at later times, which in turn may facilitate subsequent outflow episodes. 
In fact, cool gas may be removed from the core of the galaxy, leaving the central parts temporarily depleted. But the outflowing gas is likely to be bound on larger scales and remain trapped in the halo and surrounding circum-galactic medium. Indeed, recent observations indicate the presence of significant amounts of cool gas surrounding   quasar host haloes \citep{Prochaska_et_2013}. 
Several gas re-accretion events may be expected over cosmic time, with a complex combination of escape and fall-back episodes determining the subsequent evolution. 

\section{AGN variability: time dependence}
\label{Sect_AGN_variability}

The observed outflow characteristics may also be affected by the temporal variability of the central AGN. 
Indeed, AGNs are known to vary by several orders of magnitude in luminosity over a wide range of timescales. 
Radiation pressure on dust can potentially disrupt the ambient material, such that the fuel supply for the central quasar may be blown away following an outburst event.
The temporal variability of the luminosity output can have important consequences on the inferred value of the momentum flux that should be taken into account.

Observationally, the outflow is detected and its velocity measured at a certain distance from the centre, while the central luminosity is derived from the current observed flux. 
These two quantities are then used to compute the momentum ratio defined in Eq. \ref{Eq_momratio}.  
In order for the inferred values of $\zeta$ to be accurate, one requires the central luminosity to stay constant for a few million years, which is quite unlikely due to the variable nature of the AGN.
Indeed, a certain amount of time is needed for the shell to reach the current location, and over this same time span the central luminosity may not have remained the same as when the shell was initially launched. 
If e.g. the central AGN luminosity has considerably decreased over this time lapse, then the measured value of the momentum flux may be largely over-estimated.

\begin{figure}
\begin{center} 
\includegraphics[angle=0,width=0.4\textwidth]{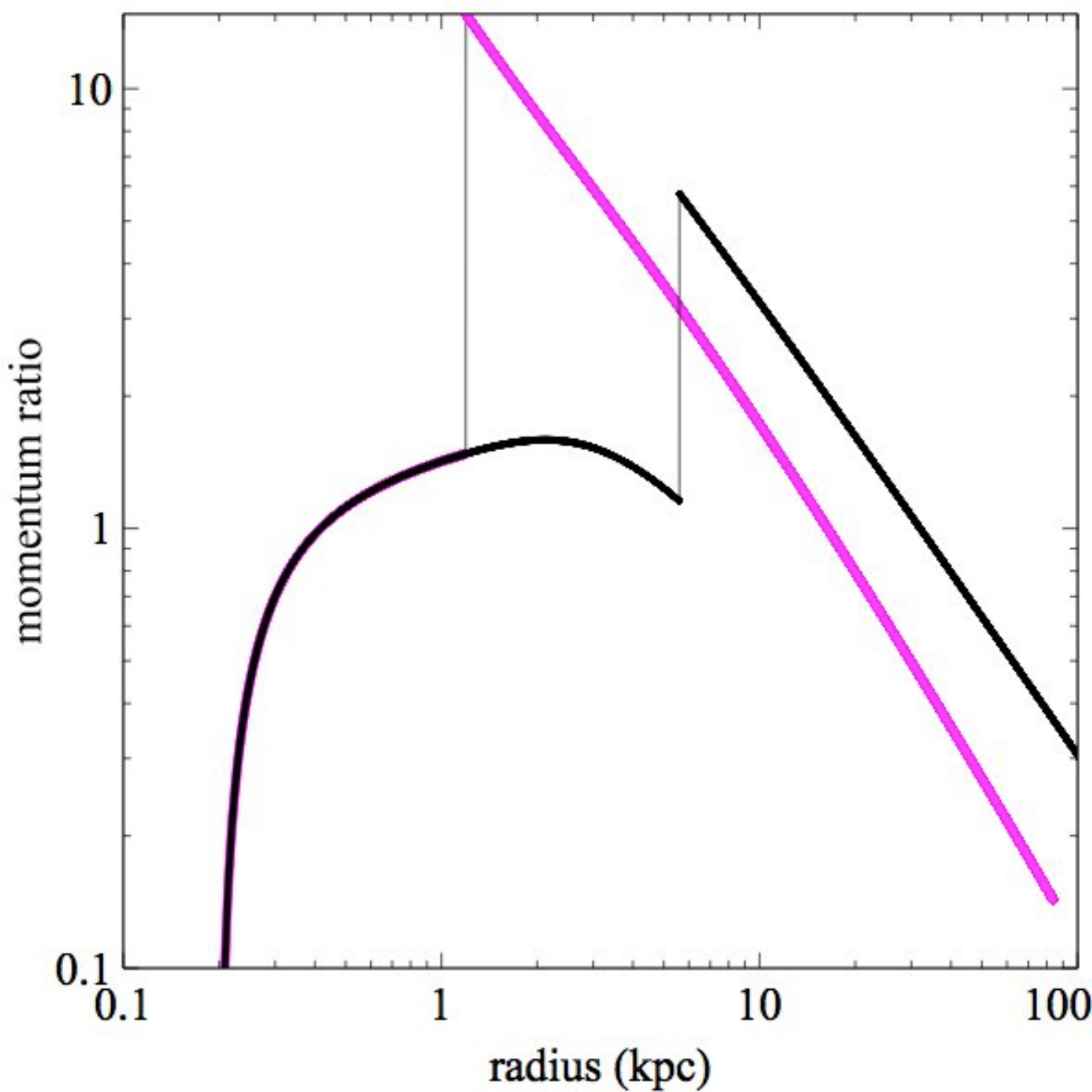} 
\caption{\small
Momentum ratio as a function of radius in the case of a sudden drop in luminosity: 
$L_1 = 5 \times 10^{46}$erg/s, $L_2 = 1 \times 10^{46}$erg/s, $t_b = 5 \times 10^{6}$yr (black); 
$L_1 = 5 \times 10^{46}$erg/s, $L_2 = 5 \times 10^{45}$erg/s, $t_b = 2 \times 10^{6}$yr (magenta). 
}
\label{Fig_AGN_drop}
\end{center}
\end{figure}

For simplicity, we first consider the case in which the central AGN undergoes a sudden drop in luminosity at a given time (Fig. \ref{Fig_AGN_drop}). 
Despite the luminosity decrease, the outflow still carries on, but an abrupt jump occurs in the the corresponding value of the momentum flux.
There are two main parameters governing the momentum ratio profile: the amplitude of the luminosity drop and the time when this break occurs. 
If the subsequent luminosity ($L_2$) is lower, then the peak value of the momentum ratio is correspondingly higher; while for a shorter $t_b$, the break occurs at a smaller radius. 
Thus, varying the two parameters, one can obtain different peak locations and distinct peak values. 

\begin{figure}
\begin{center} 
\includegraphics[angle=0,width=0.4\textwidth]{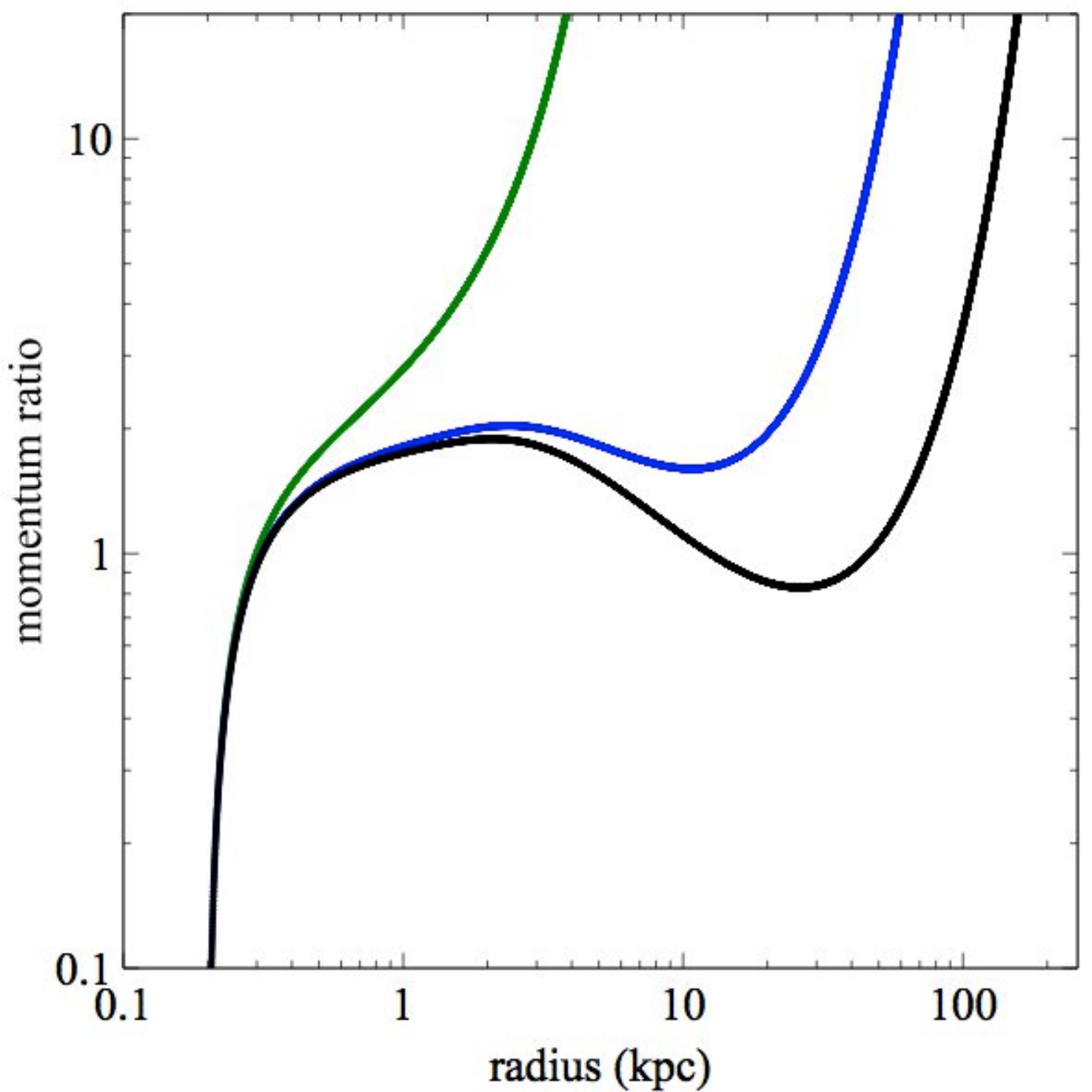} 
\caption{\small
Momentum ratio as a function of radius in the case of an exponential decay in luminosity: 
$L_0 = 10^{47}$erg/s, $t_c = 10^7$yr (black), $t_c = 5 \times 10^6$yr (blue), $t_c = 10^6$yr (green). 
}
\label{Fig_AGN_expdecay}
\end{center}
\end{figure}

We next consider the case of an exponential decay in luminosity: $L(t) = L_0 e^{-\frac{t}{t_c}}$, where $L_0$ is the initial luminosity and $t_c$ is the characteristic timescale (Fig. \ref{Fig_AGN_expdecay}).  
We see that relatively large values of the momentum flux ($\zeta \gtrsim 1$) can be reached on small scales, provided that the characteristic timescales are quite short, of the order of a few million years ($t_c \sim 10^6$yr). 
Considering that AGN accretion is limited by self-gravity (cf Discussion), the characteristic timescale is mainly determined by the mass enclosed within the self-gravity radius, and $t_c$ may be of the order of $\sim 10^5$yr at the Eddington limit \citep{King_Pringle_2007}.

The time dependence of the AGN luminosity output may be relevant in interpreting the observational results, in particular for sources accreting at sub-Eddington rates at the present time, which may form a significant fraction of the galaxy samples discussed in the literature \citep{Veilleux_et_2013, Cicone_et_2014}. 
In fact, there are several observational indications which suggest variations by a few orders of magnitude in luminosity over relatively short timescales of $\sim (10^3-10^6)$yr \citep[][and references therein]{Hickox_et_2014}. 
A nice example is given by the so-called Hanny's Voorwerp system, in which the central AGN luminosity dropped by a factor of $\gtrsim10^2$ over the last $\sim 10^5$yr, leaving the high-ionization cloud as a unique remnant of past activity \citep{Keel_et_2012}.

\section{Comparison with observed AGN outflows}
\label{Sect_Comparison}

A number of detections of massive outflows extending on galactic scales have been reported over the past few years. Observations of molecular outflows, tracing the bulk of the interstellar medium from which new stars form, are of particular relevance in determining whether the central AGN plays a role in quenching star formation. 

One of the best studied cases is Mrk 231, a nearby quasar in late stage merger, which is one of the most luminous ULIRGs in the local Universe \citep{Fischer_et_2010, Feruglio_et_2010, Rupke_Veilleux_2011}. 
Observations of OH features and CO lines indicate the presence of a massive molecular outflow, spatially resolved on $\sim$kpc scale and with an outflow rate largely exceeding the star formation rate \citep{Fischer_et_2010, Feruglio_et_2010}. 
A powerful neutral outflow, with velocities reaching $\sim 1100$km/s and extending out to radii of 2-3 kpc, is also detected in this object \citep{Rupke_Veilleux_2011}. 

Similarly, Herschel-PACS observations of local ULIRGs reveal the presence of massive molecular outflows with velocities reaching $v \sim 1000$km/s and outflow rates up to $\dot{M} \sim 1000 M_{\odot}/yr$ \citep{Sturm_et_2011}. 
The detection of molecular outflows has further been confirmed in a larger representative sample of local galaxy merger systems, including ULIRG and QSO \citep{Veilleux_et_2013}. 
Recent results based on CO observations indicate that such powerful molecular outflows commonly have high momentum fluxes, of the order of $\sim 20 L/c$ \citep{Cicone_et_2014}. 
However, one should recall that the quoted values are based on uniform spherical geometry, while in the shell limit the corresponding values should be divided by a factor of 3 \citep{Cicone_et_2014, Maiolino_et_2012}.
Moreover, one should keep in mind that the observational samples are biased toward sources with previously known outflow detections.

In our picture, large-scale, high-velocity outflows may be obtained by considering the effects of radiation pressure on dust. We have seen that radiation pressure-driven shells can reach asymptotic velocities of $\sim$1000 km/s on kpc-scales, mainly depending on the central luminosity and shell mass. 
In particular, a higher asymptotic velocity is reached for low-mass shells in high luminosity systems, with a stronger dependence on the central luminosity. 
\citet{Sturm_et_2011} find a tentative evidence for a positive correlation between terminal velocity and AGN luminosity.
A similar trend of increasing outflow velocity with increasing AGN luminosity is also reported in the larger sample of \citet{Veilleux_et_2013}, a trend in qualitative agreement with our results. 

On the other hand, in order to account for the high observed values of the momentum boosts, a number of favourable conditions need to be met.  
Following our parameter space study, the main factor governing the small-scale dynamics is the IR optical depth. 
We have seen that a high initial IR optical depth, either due to an enhanced IR opacity or a smaller initial radius, can lead to large momentum ratios. 
In fact, the highest $\zeta$ values are attained at small radii, where the medium is optically thick to the reprocessed IR emission, i.e. in the obscured phase. 
For a given IR opacity, the momentum flux is most significantly affected by variations in the initial radius: a decrease in $R_0$ can lead to a significant increase in $\zeta$ on $\lesssim$kpc scales (cf Fig. \ref{Fig_momratio_varR0}). 
In this way, one can obtain large values of the momentum flux by adjusting the initial radius, without requiring extreme values of the IR opacity. 
Based on the observed source properties and inferred measurement of the momentum flux, one can in principle derive the initial radius. 
As a specific example, we can consider the case of Mrk 231.
Adopting the reported values for the velocity dispersion ($\sigma = 120$km/s), central luminosity ($L \sim 10^{46}$erg/s), and outflowing mass ($M_{sh} \sim 3 \times 10^8 M_{\odot}$) \citep{Veilleux_et_2013, Cicone_et_2014}, we obtain values of $\zeta \lesssim 10$ at the radius of observation, consistent with the value of $\zeta = 26/3$ reported in \citet{Cicone_et_2014}, for an initial radius of $R_0 \lesssim 10$pc.
Thus the momentum flux may provide a constraint on the initial radius, requiring a quite small value of $R_0$ in this particular system (for a fiducial IR opacity). 

In principle, such method can be applied to other sources and used to infer the underlying physical quantities from the observed outflow parameters. 
However, the implementation may not be straightforward due to the uncertainties and degeneracies involved, and especially in cases where the AGN luminosity has considerably dropped since the time when the outflow was initially launched.
This may be particularly true in systems currently accreting at very sub-Eddington rates, which may comprise a significant fraction of the observational samples \citep{Veilleux_et_2013, Cicone_et_2014}. 
\citet{Veilleux_et_2013} estimate that sources in their sample accrete at sub-Eddington rates, in the range $L/L_E \sim 0.04-0.4$.
Similarly, the majority of the objects discussed in \citet{Cicone_et_2014} have rather low Eddington ratios, with some sources accreting at very sub-Eddington rates. 
For instance, Mrk 273 with an estimated black hole mass of $M_{BH} \sim 10^9 M_{\odot}$ \citep{U_et_2013} should now be accreting below $1\%$ of its Eddington limit.
As discussed in Section \ref{Sect_AGN_variability}, variations in the central AGN luminosity can seriously affect the inferred measurement of the momentum boost. Since large drops in luminosity are commonly observed \citep[][and references therein]{Hickox_et_2014}, correspondingly large values of $\zeta$ could be easily obtained. 
Observations report typical momentum rates of $\sim 10 L/c$ for molecular outflows extending on $\sim$kpc scales \citep{Cicone_et_2014}. In order to obtain high values of $\zeta$ at small radii, one requires short characteristic timescales (cf Fig. \ref{Fig_AGN_drop}). 


\section{Discussion}
\label{Sect_Discussion}

Galaxy-scale outflows, extending on $\sim$kpc scales, are now commonly detected in a number of active galaxies \citep[e.g.][]{Sturm_et_2011, Veilleux_et_2013, Cicone_et_2014}. 
The observed molecular outflows are typically characterised by high velocities ($v \sim 1000$km/s), large mass outflow rates ($\dot{M} \sim 1000 M_{\odot}/yr$), and high momentum fluxes ($\gg L/c$). 
The inferred energetics of such powerful outflows suggest that they are most likely driven by the central AGN. 
But the problem remains as how to efficiently couple the energy and momentum released by the accreting black hole to the surrounding medium. 

As mentioned in the Introduction, large-scale outflows can be driven by a number of physical mechanisms. 
For instance, two distinct regimes of outflow driving are discerned in the shocked wind scenario \citep[e.g.][]{King_et_2011}: momentum-driving at small radii ($r \lesssim 1$kpc), where the Compton cooling time is shorter than the wind flow time, and energy-driving at larger radii where the full energy of the wind is coupled to the ambient medium. 
The energetics of the observed outflows, and in particular the high measured values of the momentum ratios ($\sim 20 L/c$), seem to favour the energy driving mechanism \citep{Zubovas_King_2012, Faucher-Giguere_Quataert_2012} and apparently rule out driving by direct radiation pressure.  

However, radiation pressure on dust may be another viable mechanism for driving fast outflows on galactic scales. We note that radiation pressure acts on the bulk of the mass and do not expect developments of strong shocks in the medium. Radiative cooling of dense gas should also be efficient, allowing the dusty gas to cool down and remain cold. 
We have seen that shells driven by radiation pressure on dust can actually attain high velocities and large values of the momentum flux under certain physical conditions. 
The main parameter determining the small-scale physics and the peak of the momentum ratio is the initial IR optical depth. 
Indeed large values of the momentum flux consistent with the observed ones can be obtained, provided that the shell is initially optically thick to its own reprocessed IR radiation. 
In fact, high values of $\tau_{IR}$ are not unexpected in the central regions of dense ULIRG and obscured AGN \citep{Thompson_et_2005}. 

In general, the optical depth depends on the density and opacity of the ambient medium. 
In realistic situations, the density distribution is likely to be inhomogeneous and clumping of the interstellar medium may lead to a lowering of the global dust covering fraction. As a consequence, radiative momentum transfer may be less efficient, and the optical depth might be reduced along certain lines of sight. 
Numerical simulations of radiative feedback indicate that the radiation force on dusty gas may be reduced by a factor of $\sim2$ in cases of severe clumping compared to the case of a smooth gas distribution \citep{Roth_et_2012}. 
Although the initial launch radius may in principle be derived from the outflow parameters (Sect. \ref{Sect_Comparison}), a clumpy distribution also implies that the initial radius is not strictly defined, and the overall picture may be much more subtle.  A lower limit is only set by the dust sublimation radius, below which electron scattering dominates the local opacity.

The coupling between radiation and matter depends on the opacity of the ambient medium, which is directly proportional to the dust-to-gas ratio. An increase in the dust-to-gas fraction leads to an enhanced opacity, which in turn leads to a higher optical depth. Recent observations indicate that large amounts of dust can be produced in core-collapse supernovae and spread into the surrounding environment, implying high dust-to-gas ratios \citep{Wesson_et_2015, Owen_Barlow_2015}. Moreover, their fits to the spectral energy distribution favour large dust grains, which are also more likely to survive.  

Hydrodynamic simulations of galaxy mergers with feedback based on radiation pressure on dust require large values of the IR optical depth ($\tau_{IR} \sim 25$) in order to reproduce the observed $M-\sigma$ relation \citep{Debuhr_et_2011}. 
This suggests that much of the black hole growth occurs in the regime where matter is optically thick to the re-radiated IR emission, i.e. in the obscured phase \citep[cf][]{Fabian_1999}. 
Radiative transfer calculations, including multi-dimensional effects, indicate that winds driven by radiation pressure on dust can effectively attain momentum rates of several ($\sim 5$) L/c due to multiple scatterings \citep{Roth_et_2012}. 
On the other hand, \citet{Novak_et_2012} note that the low opacity of dust to the re-radiated IR emission and dust destruction processes (e.g. sputtering) do not allow values much exceeding $\sim L/c$. 
It has also been argued that Rayleigh-Taylor instabilities developing in the outflowing medium tend to reduce the matter-radiation coupling and thus limit the momentum transfer, such that the outflow cannot effectively exceed the single scattering limit \citep{Krumholz_Thompson_2013}. 
By contrast, the most recent results show that, despite the development of instabilities in the flow, there is a continuos acceleration of the gas, which potentially allows the propagation of the outflow on much larger scales \citep{Davis_et_2014}. 
Therefore, although several issues still remain to be resolved, radiation pressure on dust should be considered as a viable mechanism for driving large-scale outflows in AGN host galaxies. 
In fact, we have previously argued that the basic properties of galaxies, such as characteristic radius and mass, could be essentially set by the action of radiation pressure on dust \citep{Paper_3}. 

Another concern is the variability of the central AGN. 
As we have seen in Section \ref{Sect_AGN_variability}, luminosity variations can have considerable effects on the inferred value of the momentum ratio: if the current luminosity is much lower than the luminosity at the time when the shell was initially launched, then the momentum flux can be largely over-estimated. 
One may argue that AGN variability is an intrinsically random process, whereby luminosity variations occur in both directions with essentially equal probability. 
But there could also be some preferential trend, with one way being more likely than the other.  
If the outflow sweeps up and carries away large amounts of accreting matter, then potential fuel is removed from the nuclear region. As a result, the accretion rate onto the central black hole may decrease, which in turn may lead to a lower luminosity output. 
Certainly, the accretion process  is a complex phenomenon, and one needs to consider more in detail the connection with the outflow component. 

The AGN variability is also related to another interesting question regarding the outflow driving mechanism: AGN- or starburst-driven.   
When powerful outflows are observed on $\sim$kpc-scales, but no corresponding sign of AGN activity is detected at the centre, one may be led to conclude that these outflows are not AGN-driven. 
Indeed observations of high-velocity outflows from (post)-starburst galaxies have been explained in terms of compact starbursts, without the need to invoke AGN feedback \citep{Diamond-Stanic_et_2012, Sell_et_2014}. 
However, the lack of ongoing AGN activity at the present epoch does not preclude that the large-scale outflows we observe today were not driven by a past powerful AGN episode that has since faded. 
In fact, there are now indications of the intermittent and episodic nature of AGN activity, with luminosities dropping by several orders of magnitude on relatively short timescales \citep[][and references therein]{Hickox_et_2014}. 
Recent ALMA observations of an obscured quasar, presenting both compact molecular and extended ionized outflows, have been interpreted in terms of episodic AGN activity whereby the two outflow components may be driven by two distinct burst events \citep{Sun_et_2014}. 
An interesting case is the recent claim of an extended, high-velocity molecular outflow from a compact starburst \citep{Geach_et_2014}. This has been attributed to stellar feedback, as the outflow momentum flux is found to be comparable to the radiative momentum output $L/c$ from stars. 
However, if the bulk of the outflow only subtends a small solid angle at the starburst (as seems to be the case from the galaxy image, see their Fig. 3), then the comparison may break down, unless the starburst emission is particularly anisotropic. 

Given that the maximal luminosity of an AGN is likely within a factor of a few of the Eddington luminosity, a plausible limit on the momentum flux may be obtained by considering the Eddington limit of the AGN rather than the  estimate of the current luminosity.
The luminous AGN phase may last until the accretion disc empties. The actual extent of the disc is likely limited by self-gravity, with gas beyond the self-gravity radius ($R_{sg}$) being consumed in star formation rather than feeding the central black hole \citep{King_Pringle_2007}. Such accretion episodes are shown to follow a characteristic time-evolution, with the disc evolution timescale ($\tau_{sg}$) only weakly depending on the black hole mass (cf Eq. 14 in \citet{King_Pringle_2007}). 

In general, understanding the connection between the central AGN activity and star formation on galactic scales is a complex problem, and it is often not trivial to disentangle the underlying mechanism driving the observed outflows. 
In a number of situations, we may re-interpret the fast outflows attributed to central starbursts as being driven by past AGN activity.
If this is indeed the case, then AGN feedback may play an even more important role in launching galactic-scale outflows than previously thought.


\section*{Acknowledgements}

We thank Todd Thompson for detailed comments on the manuscript. 
WI acknowledges support from the Swiss National Science Foundation.

\bibliographystyle{mn2e}
\bibliography{biblio.bib}

\label{lastpage}

\end{document}